\documentclass{article}
\usepackage{spconf,amsmath,graphicx}
\usepackage{xcolor}

\usepackage{url}
\usepackage{amsfonts}
\usepackage{booktabs}
\usepackage{hyperref}
\usepackage{soul}

\def\ourmodel{PECMAE} 
\def\ourdataset{XAI-Genre} 

\title{Leveraging pre-trained autoencoders \\ for interpretable prototype learning of music audio}

\name{Pablo Alonso-Jim\'enez$^{1}$ \qquad Leonardo Pepino$^{3}$ \qquad Roser Batlle-Roca$^{1}$ \qquad Pablo Zinemanas$^{1}$}
\secondlinename{Dmitry Bogdanov$^{1}$ \qquad Xavier Serra$^{1}$ \qquad Mart\'in Rocamora$^{1,2}$}

\address{$^{1}$ Music Technology Group, Universitat Pompeu Fabra, Spain \quad $^{2}$ Universidad de la Rep\'ublica, Uruguay\\$^{3}$ Instituto de Investigación en Ciencias de la Computación (ICC), CONICET-UBA, Argentina}

\begin{document}
%
\maketitle
\begin{abstract}
We present \ourmodel\, an interpretable model for music audio classification based on prototype learning. Our model is based on a previous method, APNet, which jointly learns an autoencoder and a prototypical network. Instead, we propose to decouple both training processes. This enables us to leverage existing self-supervised autoencoders pre-trained on much larger data (EnCodecMAE), providing representations with better generalization. APNet allows prototypes' reconstruction to waveforms for interpretability relying on the nearest training data samples.
In contrast, we explore using a diffusion decoder that allows reconstruction without such dependency.
We evaluate our method on datasets for music instrument classification (Medley-Solos-DB) and genre recognition (GTZAN and a larger in-house dataset), the latter being a more challenging task not addressed with prototypical networks before.
We find that the prototype-based models preserve most of the performance achieved with the autoencoder embeddings, while the sonification of prototypes benefits understanding the behavior of the classifier.

\end{abstract}
\section{Introduction}
\label{sec:intro}
\vspace{-0.3cm}





After achieving significant breakthroughs in computer vision, speech recognition, and natural language processing, deep-learning models have become state-of-the-art in music information retrieval (MIR)~\cite{Peeters2021}. 
Yet, understanding the reasons behind the predictions of deep neural networks (DNNs) remains a challenging endeavor, motivating the increased interest in developing explanation methods and interpretable predictive models~\cite{Molnar2022}. 
%
%
%
We understand \emph{interpretability} as the capability of an algorithm or model to be comprehensible, explainable, and understandable, which allows an external observer to decipher 
its behavior and discern its decisions~\cite{AIHLEG2019}.

In the context of sound and music-related applications (such as sound engineering, music production, and music recommendation), faithful human-understandable explanations of model predictions can increase trustworthiness and enhance user experience~\cite{BarredoArrieta2020}. From a developer's perspective, an interpretable model could better reveal potential issues of its data or inner workings, allowing the detection of biases, malfunctions, or possible adversarial attacks~\cite{sturm2017horse, prinz2021end}. Ultimately, interpretability can also provide insights into the target problem, thus helping researchers learn more about it. 

%


In this paper, we introduce \ourmodel,\footnote{Prototype EnCodecMAE. 
}
a method inspired by
APNet (Audio Prototype Network)~\cite{Zinemanas2021}---an interpretable prototype-based 
audio classification model originally applied to voice commands, environmental sounds, and music instrument recognition tasks.
Prototype-based models allow interpretability by measuring similarity between model inputs and the prototypes in the encoder latent space \cite{Li2018}. Additionally, APNet features an autoencoder architecture that allows to sonify the prototypes for further insights. 
The key differences of our proposal are that we rely on a pre-trained autoencoder (EnCodecMAE~\cite{pepino2023encodecmae}) instead of jointly optimizing it for the classification task and that we use a diffusion decoder to sonify the prototypes instead of using information from specific samples of the training set. 
While the original APNet model was 
evaluated only on short consistent monophonic sounds, 
we consider music genre in addition to instrument recognition.
Music genre recognition is a more challenging task as there is no universal genre taxonomy~\cite{bogdanov2019acousticbrainz,epure2019leveraging} 
and judgments on genre encompass high-level concepts from music theory, sociohistorical context, and subjectivity~\cite{aucouturier2003representing,mckay2006musical,craft2007many,sordo2008quest}.
Still, there is a lack of experimental designs for evaluation of music genre recognition systems beyond classification accuracies, manual inspection and interpretation of features used for classification
~\cite{sturm2012survey}.

In summary, the main contributions of our work are (i) showing that it is possible to decouple the training of the autoencoder and the prototype system, which unlocks the possibility of using self-supervised autoencoders trained on larger datasets, (ii) relying on a generative 
model to eliminate the need to transfer information from specific training samples to reconstruct the prototypes, and (iii)
extending the technique of prototype-based audio classification to the task of music genre classification for the first time.





\section{Related work}
\label{sec:realated_work}

\vspace{-0.8cm}

~\subsection{Audio Prototype Network}
\vspace{-0.1cm}
Interpretability strategies in the audio domain remain scarcely explored, especially in music~\cite{batlle2023transparency}. 
Among the existing methods for sound and music classification~\cite{dileman2014, mishra2017local, won2019interpretable, loiseau22online}, only few are interpretable by design.
In particular, APNet~\cite{Zinemanas2021} is an interpretable DNN for sound classification based on prototypes that are learned during training along with the latent space encoder and decoder. The decoder is devised to reconstruct the prototypes back to the input representation (mel-spectrogram) that can be sonified. 
The model shows compelling results 
illustrating that a system can be both interpretable and accurate.
However, this model presents scalability issues regarding the number of prototypes and classes, given that the latent space is of high dimension and the prototypes are learned in this space and stored in the model. Additionally, APNet's reconstruction process transfers information on which indices were kept in the pooling layers from the encoder into the un-pooling layers of the decoder to improve the reconstruction quality. 
Assuming that the prototypes are similar to data instances, the pooling indices are extracted from the closest instance of the training set to reconstruct them, which provides interpretability suitable for end users.
However, the prototype reconstruction is strongly biased towards a training sample instead of what drives the classification decision, which would be more interesting from a developer's perspective. 
Both limitations are addressed by our proposed model.

\vspace{-0.2cm}




\subsection{EnCodecMAE}
\vspace{-0.2cm}

Neural codecs have recently emerged as a way to efficiently compress audio \cite{zeghidour2021soundstream, defossez2022high, kumar2023high}. They usually comprise a convolutional autoencoder with a Residual Vector Quantization (RVQ) layer in the bottleneck. The neural codes at the output of the RVQ layer have a low bitrate because of the 
low time resolution (75 Hz /$\sim$13 ms frames for EnCodec)
and the quantization. The high reconstruction quality of these neural codes for generic audio, together with the low frame rate, make them suitable as features for music-related tasks. However, 
because of the reconstruction objective,
these features might not contain high-level and context information spanning multiple frames of the signal.


Recently, self-supervised learning (SSL) techniques have been applied to learn higher-level and more contextualized features. EnCodecMAE \cite{pepino2023encodecmae} learns to reconstruct masked segments from the 128 dimensions of non-quantized EnCodec features using a transformer architecture. The authors showed that the resulting features, after mean-pooling over time, have a strong performance in 
a diverse set of audio-related tasks, including music genre classification (GTZAN) and pitch detection (NSynth), outperforming EnCodec and other audio embeddings. Moreover, the original waveform can be recovered from the sequence of EnCodecMAE features.

\section{Method}
\label{sec:method}


\begin{figure}[t]
    \centering
    \includegraphics[width=0.5\textwidth]{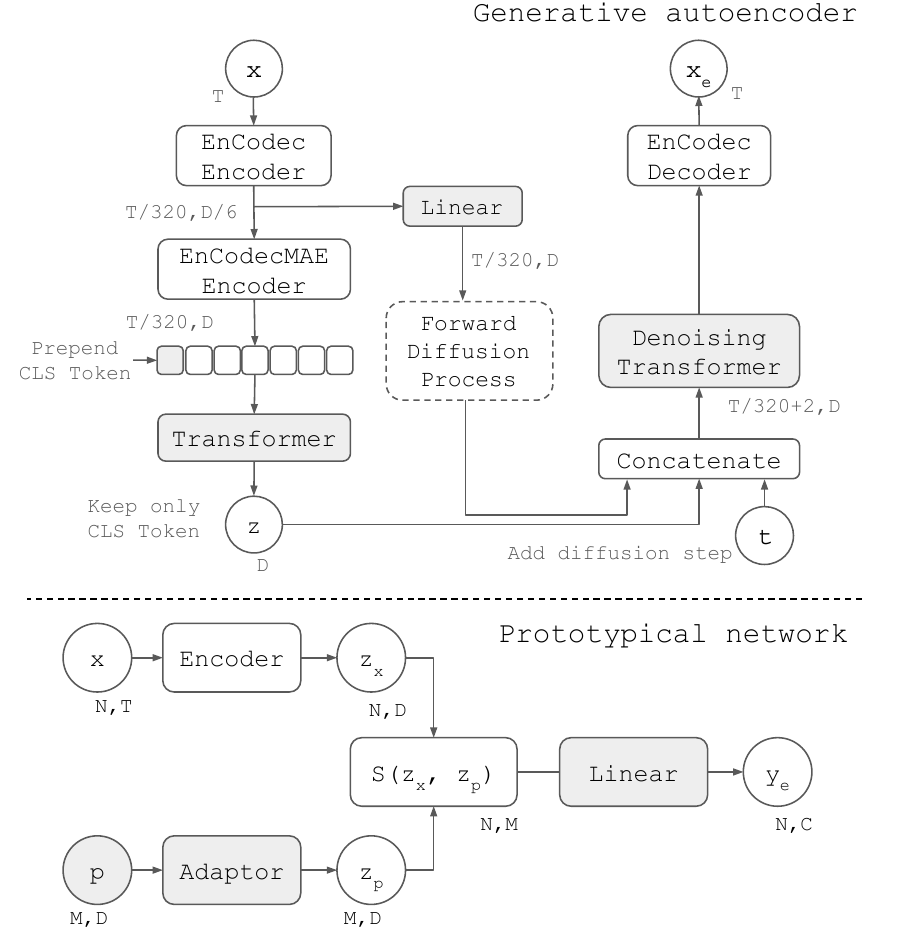}
    \caption{Diagram of the proposed \ourmodel\ model.
    The colored boxes indicate trainable modules.
    }
    \label{fig:pipeline}
\end{figure}

The main components of our system are depicted in Fig.~\ref{fig:pipeline}.
The generative autoencoder operates through the embeddings $z$. The prototype network solves a classification task by learning prototypes $p$ in the embedding space of $z$ so that they can be decoded to audio.

\vspace{-0.2cm}    
\subsection{Generative autoencoder}
\vspace{-0.2cm}
While EnCodecMAE already provides a reconstructable latent space, we observed poor prototype sonification after initial experimentation.
We hypothesized that this is because the temporal resolution is excessive and that the system would benefit from summarizing this dimension so that the prototypes were learned in a more abstract space.
To this end, we train an autoencoder on top of EnCodecMAE features, consisting of a transformer encoder that summarizes $T=4$ seconds of audio (a sequence of 300 768-dimensional EnCodecMAE features) into a single vector $z \in \mathbb{R}^{D}$ with $D=768$, and a decoder based on a latent diffusion model conditioned with $z$ to generate the EnCodec features corresponding to the original audio. We prepend a CLS token to the input of the transformer encoder and use the corresponding output element as $z$. The conditioning of the diffusion decoder is implemented by prepending $z$ to the noisy EnCodec inputs.
With this approach, our compression rate is 28 times higher than that of APNet, 
facilitating scalability both in terms of the number of prototypes and the input duration. 

\subsection{Prototypical network}
The prototypical network works by measuring the similarity, $S$, between the embeddings of a batch of input audio instances $z_x \in \mathbb{R}^{N \times D}$ and the set of prototypes $z_p \in \mathbb{R}^{M \times D}$,
where $N$ is the batch size, $M$ is the number of prototypes, $z_p$ is the projection of each prototype $p$ using an optional 1-layer transformer adaptor,\footnote{We also report results without this adaptation layer (PECMAE-NA-5).} and $S = exp^{-||z_x - z_p||^2_2}$.
Given $C$ classes, each one is assigned the same number of prototypes, $M/C$.
The prototypes corresponding to each class are initialized to centroids obtained by applying k-means clustering over the embeddings $z_x$ of the class samples. Finally, we use a linear layer to map $S$ into class logits.
This layer is initialized so that the connection of each prototype with its respective class is 1 but 0 with the others.
During training, the prototypes, the prototype transformer adaptor, and the linear layer are optimized while the autoencoder is kept frozen.

\ourmodel\ employs two loss functions. We use binary cross-entropy to optimize the classification task, $L_c$. 
In addition, we define a prototype loss that minimizes the L2 distance between each prototype and the closest sample among the instances of the same class in the batch, $z_{xc}$, 
\begin{equation}
L_p = \frac{1}{M}\sum_{j=1}^M \min_i ||z_{xc, ij} - z_{p, ij} ||^2_2 .
\end{equation}
The goal of this term is to prevent prototypes from diverging too much from real samples to favor interpretable reconstructions.
Note that we avoid using an additional loss term minimizing distances between samples and prototypes as in previous works~\cite{Zinemanas2021, Li2018} since, in our case, the sample representations are not trainable.
Finally, the losses are aggregated using a weighting factor $\lambda$, $L = \lambda L_c + (1 - \lambda) L_p$.
After training, the decoder can be used to sonify the prototypes.  

\section{Experiments and results}


We compare the classification accuracy of the proposed model with the SOTA and baseline systems and study the characteristics of learned prototypes on one music instrument and two genre classification datasets (Table~\ref{tab:datasets}).

\subsection{Datasets}

\textit{Medley-Solos-DB}~\cite{lostanlen2016deep} is an instrument recognition dataset consisting of 3-second recordings for eight instruments: clarinet, distorted electric guitar, female singer, flute, piano, tenor saxophone, trumpet, and violin.
While our main interest is in genre recognition, we considered evaluating our 
model in 
one dataset used in previous works for comparison purposes.

\textit{GTZAN} is a popular genre recognition dataset consisting of 30-second excerpts across 10 broad musical genres. We consider a filtered version of the dataset discarding duplicated and corrupted tracks identified by Sturm~\cite{sturm2013gtzan}.
To achieve a genre-balanced split, we use track IDs ending in 8 for validation (e.g., blues.00008) and in 9 for testing.

\textit{\ourdataset} is an in-house dataset of 30-second audio previews annotated by 24 genre classes, retrieved from the Spotify API,\footnote{\url{https://developer.spotify.com/documentation/web-api}} built for the purpose of this study and our planned future work on evaluation methodologies for interpretable genre recognition. 
This dataset is 20 times larger than GTZAN in terms of music tracks and includes more than twice the number of classes, adding complexity and diversity to the classification task.
\vspace{-0.2cm}

\begin{table}[]
    \small
    \centering
    \begin{tabular}{lccc}
    \toprule
    Datasets & Classes & Samples & Duration (h) \\
    \midrule
    Medley-solos-DB & 8 & 21,571 & 17.2 \\
    GTZAN & 10 & 919 & 7.6 \\
    \ourdataset & 24 & 18,634 & 155.2 \\
    \bottomrule
    \end{tabular}
    \caption{Considered datasets in terms of number of classes, samples, and total duration in hours.}
    \label{tab:datasets}
\end{table}

\subsection{Implementation details}



We train the diffusion autoencoder in the Free Music Archive dataset~\cite{fma_dataset}, composed of 106,574 30-second music tracks using batches of 128 4-second segments. The autoencoder consists of a transformer with a 2-layer encoder and an 8-layer decoder, which is then trained for 330,000 steps using the AdamW optimizer with a weight decay of 0.05 and a fixed learning rate of $1\mathrm{e}{-4}$. We apply classifier-free guidance, setting the unconditional probability to 0.1 during training, and use the V-Diffusion objective~\cite{salimans2021progressive}. Our experiments use the EnCodecMAE base model, which has 10 transformer layers and 12 attention heads and is trained in a mixture of Audioset, Librilight Medium, and Free Music Archive.

We train \ourmodel\ models in the GTZAN, Medley-Solos-DB, and \ourdataset\ datasets featuring 1 to 40 prototypes per class. In all cases, we use z-score normalization and a batch size of 256 samples. 
We prefer a rather larger batch size, considering that a larger pool of tracks to approximate our embeddings will lead to a better embedding reconstruction. 
All the models are trained for 150,000 steps using the Adam optimizer with a weight decay of $1\mathrm{e}{-5}$ using the OnceCycleLR learning rate scheduler with a peak value of $1\mathrm{e}{-3}$.
The hyperparameter $\lambda$ controls the weight of the classification and prototype loss components.
After preliminary experiments, we set $\lambda$ to 0.25 to favor prototype reconstruction.
In all cases, we use fixed training, validation, and testing splits, and test the models using the checkpoint of the step achieving the lowest validation loss.
Since our autoencoder operates on 4-second segments, in testing we average the class logits for non-overlapping segments in the track and compute the class-normalized accuracy.

\subsection{Classification Results}
Table~\ref{tab:results} presents the metrics for all compared methods on each of the considered datasets using the same splits.
We report SOTA from literature together with a Multi-Layer Perceptron (MLP) trained on SOTA audio embeddings for genre classification (MAEST)~\cite{alonso2023efficient} and the original APNet model featuring 5 prototypes per class.
Additionally, we train MLPs with EncodecMAE (ECMAE) embeddings and its summarized version (ECMAE-S), referred to as $z$ in Fig.~\ref{fig:pipeline}. Since our prototypes are learned in the ECMAE-S space, these serve as our reference for the performance ceiling.
For \ourmodel\ we consider 1, 3, 5, 10, 20, and 40 prototypes per class, plus a version without the transformer adaptation layer (NA).

The results show that ECMAE achieves lower performance compared to methods based on large supervised datasets~\cite{alonso2023efficient} in~\ourdataset, or careful feature design~\cite{anden2019joint}\ in Medley-Solos-DB.
ECMAE-S has slightly lower accuracies than ECMAE due to information loss associated with the higher compression rate but produces better sonification results.\footnote{Operating in ECMAE's higher-dimensional space resulted in poor sonification of the prototypes. 
}
As a general trend, \ourmodel\ performance increases with the number of prototypes and is comparable to or slightly below ECMAE-S. %
Finally, we find that our method achieves higher classification performance than APNet in ~\ourdataset\ and Medley-Solos-DB, even when fewer prototypes per class are used.
We hypothesize that this is due to the powerful representations of EncodecMAE, which had been trained in a much larger data collection and already showed good performance in music-related tasks~\cite{pepino2023encodecmae}.



\begin{table}[]
    \centering
    \small
    \addtolength{\tabcolsep}{-2pt}
    \begin{tabular}{lcccc}
    \toprule
          & Params. & GTZAN & Medley & \ourdataset  \\
    \midrule
    \small \textcolor{gray}{\textit{State of the Art}} \\
    
    Literature &   & 82.1~\cite{kim2018sample} & 78.0~\cite{anden2019joint} & - \\ 
    
    MLP MAEST & 300K & 95.6 & - & 62.9 \\
    \midrule
    \small \textcolor{gray}{\textit{Baseline}} \\
    APNet-5 & 2.7-4.2M & 87.4 & 65.8 & 48.0 \\
    \midrule
    \small \textcolor{gray}{\textit{Ceiling}} \\
    MLP ECMAE & 100K & 85.7 & 75.7 & 58.0 \\
    MLP ECMAE-S & 100K & 85.9 & 72.1 & 56.1 \\
    \midrule
    \small \textcolor{gray}{\textit{Ours}} \\
    \ourmodel-NA-5 & 31-90K & 81.8 & 66.8 & 44.0 \\
    \ourmodel-1 & 5.5M & 80.8 & 63.9 & 44.0 \\
    \ourmodel-3 & 5.6M & 82.8 & 67.6 & 48.6 \\
    \ourmodel-5 & 5.6M & 83.8 & 70.2 & 50.1 \\
    \ourmodel-10 & 5.6M & \textbf{86.9} & 71.1 & 51.8 \\
    \ourmodel-20 & 5.6M & 85.9 & 69.2 & 52.8 \\
    \ourmodel-40 & 5.8M & 85.7 & \textbf{71.3} & \textbf{53.6} \\
    \bottomrule
    
    \end{tabular}
    \caption{Normalized classification accuracies. 
    }
    \label{tab:results}
\end{table}

\subsection{Effect of the decoder}
\vspace{-0.2cm}

Since our autoencoder relies on a generative model, its decoder is constrained to synthesize audio close to its training distribution, which can result in reconstruction biases.
After preliminary tests, we found that a decoder based on V-diffusion was providing more faithful reconstructions than an alternative based on a conditional language model over EnCodec tokens decoder using a GPT2 architecture.
We verified that the important class information was not altered in the decoding process by measuring the class predictions for the synthesized prototypes (above 99\% accuracy for ~\ourmodel-20).


    
\vspace{-0.3cm}
\subsection{Sonifying the prototypes}
\vspace{-0.1cm}
As part of developing the proposed models, we conducted iterative listening examinations of the synthesized prototypes.\footnote{Prototype sonifications and complementary results available at \\ \url{https://palonso.github.io/pecmae/}}
We observe that the sonification of the prototypes results in sounds that are far from resembling real class instances and instead have a sonic texture quality. However, we can identify many of the classes both in the case of instruments and music genres from blind listening to the prototypes. Clearly, the sonification of instrument prototypes is more convincing than genre sonification due to lower sound complexity (monophonic notes vs complete full-mix music tracks). Increasing the number of prototypes tends to provide similar-sounding prototypes, even though we can identify differences in some cases (e.g., pitches or types of timbre).
%
Notably, the autoencoder was able to reconstruct instances from all classes in our datasets with much better quality than the prototypes, which suggests that the bottleneck is in the statistical averaging process when learning the prototypes and not in the decoder itself. 
Overall, the sonification would not be appropriate for end users but is insightful for model developers \cite{sturm2017horse}, especially revealing how adversarial attacks can be devised~\cite{prinz2021end}.



%



\vspace{-0.3cm}

\section{Conclusions and future work}

\vspace{-0.3cm}

We propose an interpretable classification system that learns prototypes in the embedding space of an autoencoder and enables their sonification.
Our results show that it is possible to achieve
prototype-based models that do not notably degrade classification performance while providing a certain level of interpretability, which proves helpful in developing classification models. 
Through sonification, we observe that the models learn sonic textures instead of more complex temporal structures as the basis of their classification decisions for most of the classes.
The proposed approach motivates new directions for interpretable music audio models.
We will consider replacing the self-supervised embeddings with representations optimized for the classification tasks and investigate auto-encoding techniques that 
handle longer input sequences and increase the diversity of the learned prototypes.\footnote{This work has been supported by the Musical AI project - PID2019-111403GB-I00/AEI/10.13039/501100011033, funded by the Spanish Ministerio de Ciencia e Innovación and the Agencia Estatal de Investigación.}






\bibliographystyle{IEEEbib}
{
\small
\bibliography{refs}

\begin{thebibliography}{10}

\bibitem{Peeters2021}
Geoffroy Peeters,
\newblock ``The deep learning revolution in {MIR}: The pros and cons, the needs and the challenges,''
\newblock in {\em Perception, Representations, Image, Sound, Music}, 2021.

\bibitem{Molnar2022}
Christoph Molnar,
\newblock {\em Interpretable Machine Learning},
\newblock 2022.

\bibitem{AIHLEG2019}
{AI HLEG},
\newblock ``{Ethics Guidelines for Trustworthy AI},'' 2019.

\bibitem{BarredoArrieta2020}
Alejandro~Barredo Arrieta, Natalia Díaz-Rodríguez, Javier~Del Ser, Adrien Bennetot, Siham Tabik, Alberto Barbado, Salvador Garcia, Sergio Gil-Lopez, Daniel Molina, Richard Benjamins, Raja Chatila, and Francisco Herrera,
\newblock ``Explainable artificial intelligence ({XAI}): Concepts, taxonomies, opportunities and challenges toward responsible {AI},''
\newblock {\em Information Fusion}, vol. 58, 2020.

\bibitem{sturm2017horse}
Bob~L. Sturm,
\newblock ``The “horse” inside: Seeking causes behind the behaviors of music content analysis systems,''
\newblock {\em Comput. Entertain.}, vol. 14, no. 2, jan 2017.

\bibitem{prinz2021end}
Katharina Prinz, Arthur Flexer, and Gerhard Widmer,
\newblock ``On end-to-end white-box adversarial attacks in music information retrieval,''
\newblock {\em Transactions of the International Society for Music Information Retrieval}, vol. 4, no. 1, 2021.

\bibitem{Zinemanas2021}
Pablo Zinemanas, Martín Rocamora, Marius Miron, Frederic Font, and Xavier Serra,
\newblock ``An interpretable deep learning model for automatic sound classification,''
\newblock {\em Electronics}, vol. 10, no. 7, 2021.

\bibitem{Li2018}
Oscar Li, Hao Liu, Chaofan Chen, and Cynthia Rudin,
\newblock ``Deep learning for case-based reasoning through prototypes: A neural network that explains its predictions,''
\newblock in {\em The 32nd AAAI Conf. on Artificial Intelligence}, 2018.

\bibitem{pepino2023encodecmae}
Leonardo Pepino, Pablo Riera, and Luciana Ferrer,
\newblock ``{EnCodecMAE}: Leveraging neural codecs for universal audio representation learning,''
\newblock {\em preprint arXiv:2309.07391}, 2023.

\bibitem{bogdanov2019acousticbrainz}
Dmitry Bogdanov, Alastair Porter, Hendrik Schreiber, Juli{\'a}n Urbano, and Sergio Oramas,
\newblock ``The acousticbrainz genre dataset: Multi-source, multi-level, multi-label, and large-scale,''
\newblock in {\em Intl. Society for Music Information Retrieval (ISMIR)}, 2019.

\bibitem{epure2019leveraging}
Elena~V Epure, Anis Khlif, and Romain Hennequin,
\newblock ``Leveraging knowledge bases and parallel annotations for music genre translation,''
\newblock in {\em Intl. Society for Music Information Retrieval Conf. (ISMIR)}, 2019.

\bibitem{aucouturier2003representing}
Jean-Julien Aucouturier and Francois Pachet,
\newblock ``Representing musical genre: A state of the art,''
\newblock {\em Journal of new music research}, vol. 32, no. 1, 2003.

\bibitem{mckay2006musical}
Cory McKay and Ichiro Fujinaga,
\newblock ``Musical genre classification: Is it worth pursuing and how can it be improved?,''
\newblock in {\em Intl. Society for Music Information Retrieval (ISMIR)}, 2006.

\bibitem{craft2007many}
Alastair J.~D. Craft, Geraint~A. Wiggins, and Tim Crawford,
\newblock ``How many beans make five? the consensus problem in music-genre classification and a new evaluation method for single-genre categorisation sysytems,''
\newblock in {\em Intl. Conf. on Music Information Retrieval (ISMIR)}, 2007.

\bibitem{sordo2008quest}
Mohamed Sordo, Oscar Celma, Martin Blech, and Enric Guaus,
\newblock ``The quest for musical genres: Do the experts and the wisdom of crowds agree?,''
\newblock in {\em Intl. Conf. on Music Information Retrieval (ISMIR)}, 2008.

\bibitem{sturm2012survey}
Bob~L. Sturm,
\newblock ``A survey of evaluation in music genre recognition,''
\newblock in {\em Intl. Workshop on Adaptive Multimedia Retrieval}, 2012.

\bibitem{batlle2023transparency}
Roser Batlle-Roca, Emila G{\'o}mez, WeiHsiang Liao, Xavier Serra, and Yuki Mitsufuji,
\newblock ``Transparency in music-generative {AI}: A systematic literature review,''
\newblock {\em preprint (under reivew)}, 2023.

\bibitem{dileman2014}
Sander Dieleman and Benjamin Schrauwen,
\newblock ``End-to-end learning for music audio,''
\newblock in {\em IEEE Intl. Conf. on Acoustics, Speech and Signal Processing (ICASSP)}, 2014.

\bibitem{mishra2017local}
Saumitra Mishra, Bob~L Sturm, and Simon Dixon,
\newblock ``Local interpretable model-agnostic explanations for music content analysis.,''
\newblock in {\em Intl. Society for Music Information Retrieval Conf. (ISMIR)}, 2017.

\bibitem{won2019interpretable}
Minz Won, Sanghyuk Chun, and Xavier Serra,
\newblock ``Toward interpretable music tagging with self-attention,''
\newblock {\em preprint arXiv:1906.04972}, 2019.

\bibitem{loiseau22online}
Romain Loiseau, Baptiste Bouvier, Yan Teytaut, Elliot Vincent, Mathieu Aubry, and Loic Landrieu,
\newblock ``A model you can hear: Audio identification with playable prototypes,''
\newblock in {\em Intl. Conf. on Music Information Retrieval (ISMIR)}, 2022.

\bibitem{zeghidour2021soundstream}
Neil Zeghidour, Alejandro Luebs, Ahmed Omran, Jan Skoglund, and Marco Tagliasacchi,
\newblock ``{SoundStream}: An end-to-end neural audio codec,''
\newblock {\em IEEE/ACM Transactions on Audio, Speech, and Language Processing}, vol. 30, 2021.

\bibitem{defossez2022high}
Alexandre D{\'e}fossez, Jade Copet, Gabriel Synnaeve, and Yossi Adi,
\newblock ``High fidelity neural audio compression,''
\newblock {\em preprint arXiv:2210.13438}, 2022.

\bibitem{kumar2023high}
Rithesh Kumar, Prem Seetharaman, Alejandro Luebs, Ishaan Kumar, and Kundan Kumar,
\newblock ``High-fidelity audio compression with improved {RVQGAN},''
\newblock {\em preprint arXiv:2306.06546}, 2023.

\bibitem{lostanlen2016deep}
Vincent Lostanlen and Carmine-Emanuele Cella,
\newblock ``Deep convolutional networks on the pitch spiral for musical instrument recognition,''
\newblock in {\em Intl. Society for Music Information Retrieval Conf. (ISMIR)}, 2016.

\bibitem{sturm2013gtzan}
Bob~L. Sturm,
\newblock ``The {GTZAN} dataset: Its contents, its faults, their effects on evaluation, and its future use,''
\newblock {\em preprint arXiv:1306.1461}, 2013.

\bibitem{fma_dataset}
Micha\"el Defferrard, Kirell Benzi, Pierre Vandergheynst, and Xavier Bresson,
\newblock ``{FMA}: A dataset for music analysis,''
\newblock in {\em Intl. Society for Music Information Retrieval Conf. (ISMIR)}, 2017.

\bibitem{salimans2021progressive}
Tim Salimans and Jonathan Ho,
\newblock ``Progressive distillation for fast sampling of diffusion models,''
\newblock in {\em Intl. Conf. on Learning Representations (ICLR)}, 2021.

\bibitem{alonso2023efficient}
Pablo Alonso-Jim{\'e}nez, Xavier Serra, and Dmitry Bogdanov,
\newblock ``Efficient supervised training of audio transformers for music representation learning,''
\newblock in {\em Intl. Society for Music Information Retrieval Conf. (ISMIR)}, 2023.

\bibitem{anden2019joint}
Joakim And{\'e}n, Vincent Lostanlen, and St{\'e}phane Mallat,
\newblock ``Joint time--frequency scattering,''
\newblock {\em IEEE Transactions on Signal Processing}, 2019.

\bibitem{kim2018sample}
Taejun Kim, Jongpil Lee, and Juhan Nam,
\newblock ``Sample-level {CNN} architectures for music auto-tagging using raw waveforms,''
\newblock in {\em IEEE Intl. Conf. on Acoustics, Speech and Signal Processing (ICASSP)}, 2018.

\end{thebibliography}
}

\end{document}